\documentclass[reprint, aps,floatfix,showpacs,superscriptaddress]{revtex4-1}
\usepackage{graphicx,epsf,epsfig,epstopdf}
\usepackage{graphicx}
\usepackage{subfigure}

\begin{document}
%
\title{A tunnel FET compact model including non-idealities with verilogy implementation}

\author{Redwan N. Sajjad}
\affiliation{Microsystems Technology Laboratories, Massachusetts Institute of Technology, Cambridge, MA-02139.}
\author{Ujwal Radhakrishna} 
\affiliation{Microsystems Technology Laboratories, Massachusetts Institute of Technology, Cambridge, MA-02139.}
\author{Dimitri A. Antoniadis}
\affiliation{Microsystems Technology Laboratories, Massachusetts Institute of Technology, Cambridge, MA-02139.}


\begin{abstract}
We present a compact model for Tunnel Field Effect Transistors (TFET), that captures several non-idealities such as the Trap Assisted Tunneling (TAT) originating from interface traps ($D_\mathrm{it}$), along with Verilog-A implementation. We show that the TAT, together with band edge non-abruptness known as the Urbach tail, sets the lower limit of the sub-threshold swing and the minimum achievable current at a given temperature. Presence of charged trap states also contributes to reduced gate efficiency. We show that we can decouple the contribution of each of these processes and extract the intrinsic sub-threshold swing from a given experimental data. We derive closed form expressions of channel potential, electric field and effective tunnel energy window to accurately capture the essential device physics of TFETs. We test the model against recently published experimental data, and simulate simple TFET circuits using the Verilog-A model. The compact model provides a framework for TFET technology projections with improved device metrics such as better electrostatic design, reduced TAT, material with better transport properties etc. 
\end{abstract}

\maketitle
\section{Introduction}
Tunnel Field Effect Transistors are promising candidates for low power logic applications \cite{seabaugh2010}. They have the potential to reduce energy dissipation by relying on Band To Band Tunneling (BTBT) for carrier injection,  achieve steep turn-ON and thus reduce the supply voltage. Under ideal conditions, device simulations consistently reported switching at sub-thermal rates \cite{bhuwalka2005simulation,nayfeh2008design,bowonder2008low, khatami2009steep, luisier2009atomistic, koswatta2010possibility,avci2013energy}. Even though the output current from TFET may be low, a sub-thermal sub-threshold swing has the potential to drastically reduce power dissipation and therefore it is attractive for low power applications. Many compact models have been developed in the past to facilitate circuit simulation in Spice \cite{hong2009spice,mojumder2009band,zhang2012analytical,zhang2012compact,tanaka2016implementation}. However, in most cases the models are tested against results from device simulators instead of experimental devices \cite{hong2009spice,yang2010tunneling,zhang2012analytical,bhushan2012dc,pan2012quasi,lu2015,lu2016universal}. This is likely due to the fact that most experimental results deviate substantially from ideal device simulations and do not produce sub-thermal switching behavior. Therefore there is a clear disconnect between experimental results and compact models and the realistic potential of TFET based circuits is still unknown. 

In this paper, we address this disconnect by developing a physics based compact model that 1) fits experimental data, 2) explains the physics of non-idealities with compact expressions and 3) describes circuit performance for different levels of non-idealities and therefore lays a pathway for studying low power circuits based on TFETs. The main reasons of non-ideal switching behavior in TFETs is the existence of interface traps and non-abrupt density of states (Urbach tail) at the band edges. In the past we developed numerical models explaining the impact of trap assisted tunneling (TAT) \cite{sajjad2016trap} and how, combined with non-abrupt Urbach tail, it increases the sub-threshold swing of TFETs. In this paper, we present compact expressions of TAT based on Shockley-Read-Hall formalism. We also present simple expressions of channel potential and electric field and explain how they capture the details of TFET device physics such as the TFET quantum capacitance, super-linear output current and current saturation mechanism. We implement the model in Verilog-A and present TFET based inverter and oscillator circuit performance based on  existing TFET data. The paper is organized as follows: Section II gives a brief overview of the impacts of TAT and the compact expressions of current from TAT. Section III describes the electrostatic model and Section IV describes the BTBT model that includes the Urbach tail. In Section V, we provide model fits to experimental data. The final Section (VI) describes the circuit simulation results using the compact model in Verilog-A. 

\section{Trap assisted tunneling model}
\begin{figure}
\centering
\includegraphics[width=3.4in]{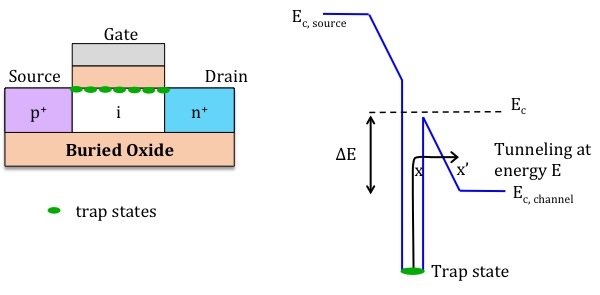}
\caption{(a) We consider the interface traps between the gate oxide and the channel. Although traps exist throughout the interface, substantial trap assisted tunneling (TAT) takes place only in the region where the electric field is high (source-channel junction). TAT can be of any combination of thermal emission (vertical transition along energy axis) and tunneling (horizontal spatial transition) as shown. 
}
\label{tat}
\end{figure}
Fig. \ref{tat} shows the device structure of a top-gate TFET and the position of the traps. The trap assisted tunneling is only strong where the electric field is high, in this case the source-channel junction. Below the threshold voltage, $V_t$ (when BTBT is triggered), electron excitation by a phonon from the valence band to a trap state followed by tunneling into the conduction band (Fig. \ref{tat}b) can give rise to leakage current. The problem is to find how much is the leakage floor compared to the BTBT current and how the overall sub-threshold swing is affected. 
From the SRH formalism, the electron generation rate from a trap to the conduction band can be written as, $e_{n0} = \frac{1}{\sigma v_\mathrm{th}D_\mathrm{it}}$, where $\sigma$ is the carrier capture cross section, $v_\mathrm{th}$ is the thermal velocity and $D_\mathrm{it}$ is the density of traps. Under high electric field, we have substantial band bending and in addition to thermal emission, electrons can partially be excited by phonons and then tunnel into the conduction band via tunneling. The overall process can be quantified by the original transition rate times an enhancement factor due to tunneling. The new rate becomes, $e_{n} = e_{n0}\times\Gamma$, where $\Gamma$ accounts for the tunneling process. At energy $E$, the emission probability is enhanced since electrons are emitted into a lower energy level ($E_c$) than they would normally emit to. The emission rate is therefore enhanced to $e_{n0}\mathrm{exp}({(E_c-E)/(k_BT)})$ \cite{furlan2001}. Accounting for the transmission probability ($Tr$) through the triangular barrier (from $x$ to $x^{'}$) and integrating over the energy range $\Delta E$,  $\Gamma$ is calculated as \cite{hurkx1992},
\begin{eqnarray}\label{eq:gamma}
\Gamma_\mathrm{n,p} (x) = \frac{1}{k_BT}\int_{E_c-\Delta E_\mathrm{n,p}(x)}^{E_c}\mathrm{exp}(\frac{E_c-E}{k_BT})Tr(E)dE
\end{eqnarray}$E$ is the energy to which the electron (or hole) is tunneling to (Fig. \ref{tat}b).  It can be shown,
\begin{eqnarray}\label{eq:gamma2}
\Gamma_\mathrm{n,p}(x) &=& \frac{\Delta E_\mathrm{n,p}(x)}{k_BT}\int^1_0\mathrm{exp}[\frac{\Delta E_\mathrm{n,p}(x)}{k_BT}u-K_{n,p}u^{3/2}]\mathrm{du}\nonumber\\
\end{eqnarray}where $K_\mathrm{n,p} = \frac{4}{3}\frac{\sqrt{2m_\mathrm{n,p}^*\Delta E_\mathrm{n,p}^3}}{q\hbar \mathcal{E}}$,
$\Delta E_\mathrm{n,p}$ defines the range of energy to which the electron (or hole) can tunnel to and from the trap. The subscript (n,p) indicates that the equations are equally applicable to both electron and holes.
\begin{figure}
\centering
\includegraphics[width=3.0in]{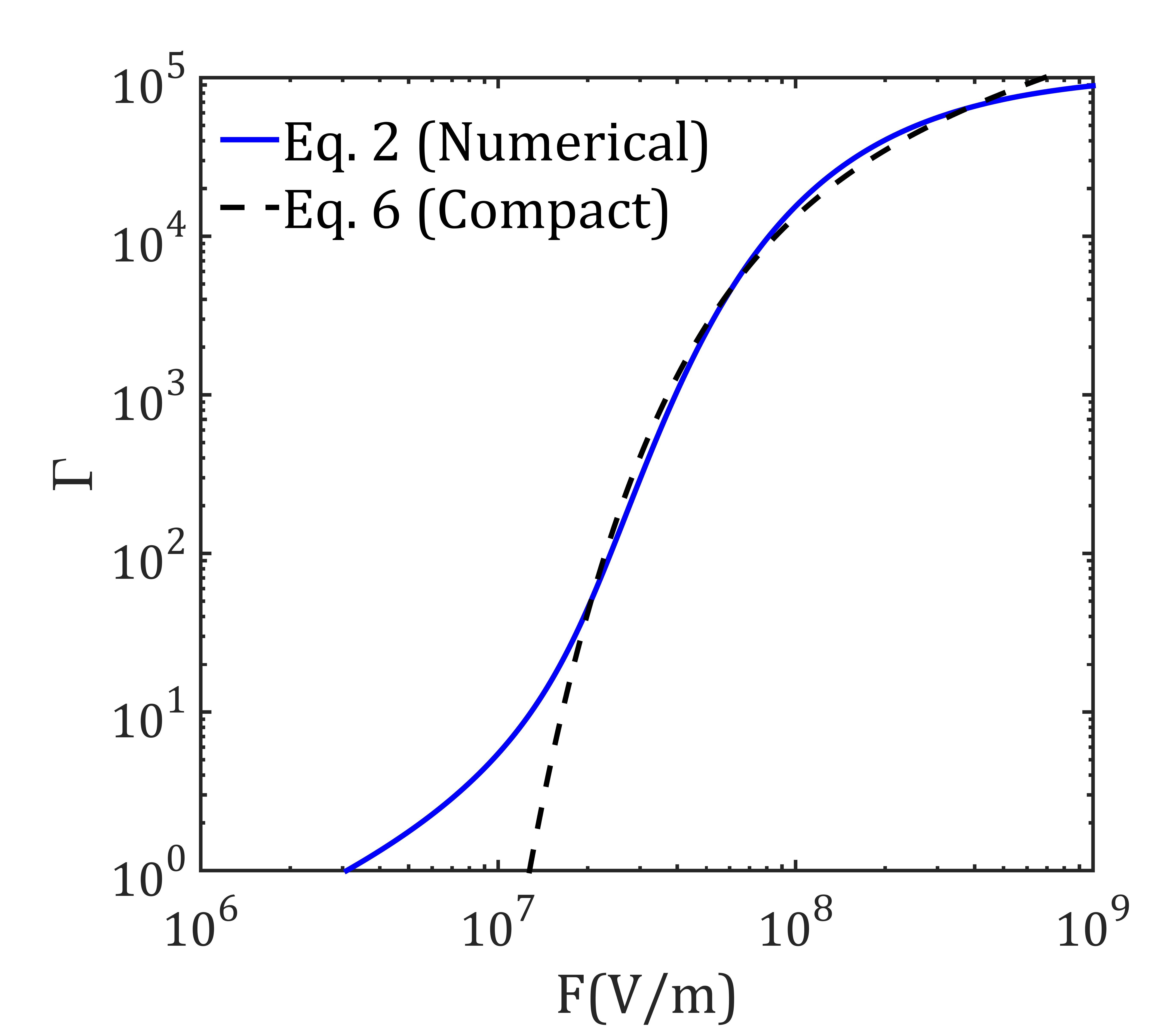}
\caption{$\Gamma$ represents the enhancement in carrier generation (compared to classical SRH formalism) due to tunneling. Since tunneling is electric field dependent, $\Gamma$ changes with the electric field. This figure shows the accuracy of the compact expression of $\Gamma$ (Eq. \ref{gamma3}) vs. the exact numerical calculation (Eq. \ref{eq:gamma2})}
\label{gamma}
\end{figure}

Once $\Gamma$ is calculated, the current can be obtained from the net generation rate following the same method as the conventional SRH formalism, 
\begin{eqnarray}\label{eq:srh}
G^n(x) &=& \int\frac{n_i^2-np}{\tau_p\frac{n+n_1}{1+\Gamma_p(x)}+\tau_n\frac{p+p_1}{1+\Gamma_n(x)}}D_\mathrm{it}\,dE\\
I &=& qW\int G^n(x)\,{dx}
\end{eqnarray}where $n_i$ is the intrinsic carrier concentration, $n$ and $p$ are the electron and hole densities. With $\Gamma=0$, Eq. \ref{eq:srh} reduces to the conventional SRH formalism. 
\subsection{Compact expression of TAT} 
Eq. \ref{eq:gamma2} involves integration and therefore is not suitable for circuit simulation. Assuming that lifetimes and electric field enhancement factors are the same for electron and hole, and that the channel is depleted of free carriers ($n=p=0$) near the source-channel junction where most of the generation takes place, and constant generation rate over a length $d_\mathrm{gen}$, we can simplify the above formalism into a compact form as below,
\begin{eqnarray}\label{compact_current}
I_{TAT} = qW\frac{n_i}{2\tau}\Gamma d_\mathrm{gen}\Big[1-e^{-\frac{qV_\mathrm{DS}}{k_BT}}\Big]
\end{eqnarray} The integration over energy is eliminated because only the midgap traps contribute significantly to the TAT. Typical channel electric field in TFETs vary around 1 MV/cm. In this high electric field regime ($K<2/3\Delta E/k_BT$), the $\Gamma$ expression can be simplified as following,
\begin{eqnarray}\label{gamma3}
\Gamma = \frac{\Delta E}{k_BT}\sqrt{\frac{2\pi}{3K}}\mathcal{F}e^{\Big[\frac{\Delta E}{k_BT}-K\Big]}\end{eqnarray}where $\mathcal{F}$ is a fitting parameter. Thus $\Gamma$ depends on the temperature $T$, electric field $\mathcal{E}$, and the material parameters effective mass ($m^*$), bandgap ($E_g$). Fig. \ref{gamma} shows the comparison of the calculation of $\Gamma$ from Eqs. \ref{eq:gamma2} and \ref{gamma3} at room temperature, assuming $\Delta E = 0.4$ eV, $m^*$ = 0.04$m_0$. The compact expression (Eq. \ref{gamma3}) shows good agreement with the exact numerical calculation (Eq. \ref{eq:gamma2}) above $\mathcal{E}=2\times10^7$ V/m with $\mathcal{F}$ = 2. However, $\Gamma$ in Eq. \ref{gamma3} represents an average enhancement factor over the entire generation volume (where the TAT takes place) and thus $\Delta E$ loses its direct physical meaning. 

\section{Electrostatics model} 
\begin{figure}
\centering
\includegraphics[width=2.7in]{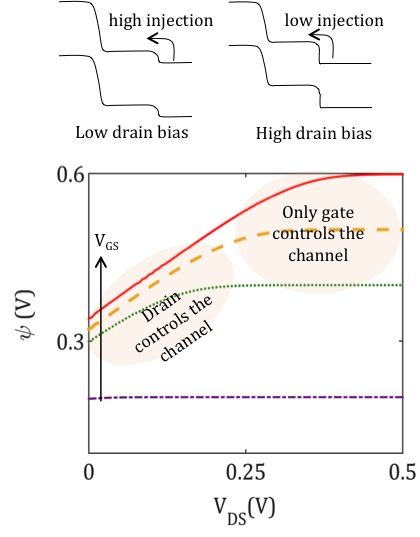}
\caption{In a TFET the channel is populated only by the injection from the drain. For low drain bias the drain-channel barrier is small therefore the resulting high carrier injection and quantum capacitance pins the electrostatic potential of the channel. At high drain bias, the injection from drain is low and the channel potential changes according to the gate voltage.}
\label{drain_control}
\end{figure}
\begin{figure}
\centering
\includegraphics[width=3in]{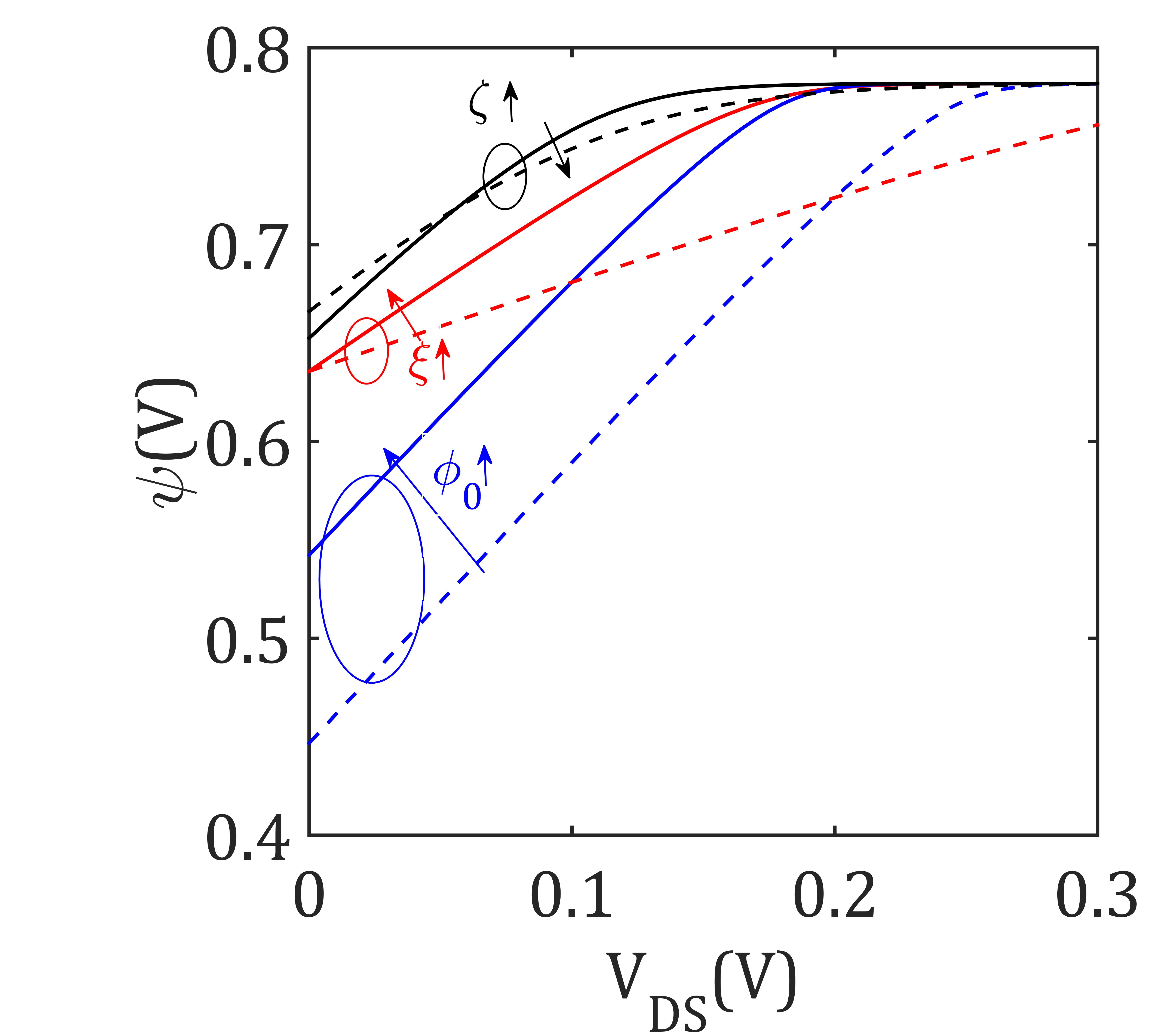}
\caption{Depending on the magnitude of the quantum capacitance, the channel potential changes with drain bias. The quantum capacitance parameters ($\phi$ and $\xi$) are used to capture the change of $\psi$ with $V_\mathrm{DS}$. $\phi_0$ is the zero bias surface potential, $\xi$ sets the rate of change of $\psi$ with $V_\mathrm{DS}$ and $\zeta$ changes the smoothness of $\psi$.}
\label{psi}
\end{figure}

The TFET surface potential is strongly influenced by the drain voltage since the channel is primarily populated with drain injected carriers. At low drain bias, the carrier injection is high and therefore the channel potential is pinned due to high quantum capacitance.  At the high drain bias limit, the injection is low and the potential is controlled primarily by the gate voltage (Fig. \ref{drain_control}). We capture the surface potential including these effects with the following empirical compact expression,
\begin{eqnarray}\label{eq:psi}
\psi = \frac{k_BT}{\zeta}\Big\{\log\Big[\log\{1+\mathrm{exp}\Big(\frac{V_\mathrm{GS,internal}-\phi}{k_BT/\zeta}\Big)\}\Big]+...\nonumber\\\frac{\phi}{k_BT/\zeta}\Big\}\nonumber\\
\end{eqnarray}where $\phi= \phi_0+\xi V_\mathrm{DS}$ and $\phi_0$ sets the zero bias ($V_\mathrm{DS}=0$) surface potential. $\xi$ sets the rate of change of surface potential with $V_\mathrm{DS}$. 
We show that this is a powerful expression that can capture the details of TFET transport, such as the transconductance, output resistance, super-linear drain current at low $V_\mathrm{DS}$ etc.  The fitting parameters for the potential are the zero-bias maximum potential $\phi_0$, the smoothness parameter $\zeta$ and the bias control parameter $\xi$. For small gate voltage ($V_\mathrm{GS}<<\phi$), Eq. \ref{eq:psi} gives , $\psi \approx k_BT/\zeta \{\log[\exp(V_\mathrm{GS,internal}-\phi)/(k_BT/\zeta)]+\phi/(k_BT/\zeta)\} \approx V_\mathrm{GS}$. For $V_\mathrm{GS}>>\phi$, we find  $\psi \approx k_BT/\zeta \{\log[(V_\mathrm{GS, internal}-\phi)/(k_BT/\zeta)]+\phi/(k_BT/\zeta)\} \approx \phi$. Therefore $\phi_0$ essentially sets the maximum potential that can be achieved in the device at low drain bias (high carrier injection). If the device is initially in $p$-$i$-$n$ regime, $\phi_0$ is roughly equal to half of the bandgap. 
Such drain bias dependence of the channel potential does not originate from any short channel effects and it is intrinsic to TFET. The parameter $\xi$ reflects the control of drain bias on the channel potential. The surface potential saturates roughly at $\phi_0$ plus the drain bias ($\xi$=1), however, materials with low quantum capacitance will have lower values of $\xi$. Fig. \ref{psi} shows how the channel potential varies with drain bias for different values of the fitting parameters used in $\psi$. For instance, as the smoothness parameter $\zeta$ is increased, $\psi$ changes more slowly and smoothly. For smaller and smaller values of $\xi$, $\psi$ changes more slowly with the drain bias. Finally, as $\phi_0$ is increased, the initial value (zero bias) of $\psi$ increases. We will later show how these parameters affect the current-voltage characteristics. 

The electric field in the channel can be expressed as,
\begin{eqnarray}
|\mathcal{E}| = \mathcal{E}_0+\frac{\psi}{\lambda}
\end{eqnarray} where $\mathcal{E}_0$ is the initial electric field (at zero gate voltage, set to $E_g/2\lambda$ for an intrinsic channel) and $\lambda$ is the characteristic scaling length, which is a function of semiconductor and oxide thicknesses. The internal gate voltage is found ($V_\mathrm{GS,internal} = \eta_tV_\mathrm{GS}$) after accounting for the gate efficiency limited by charged traps,
\begin{eqnarray}
\eta_t &=& \frac{C_\mathrm{ox}}{C_\mathrm{ox}+C_\mathrm{it}}\\
C_{it} &=& q^2\int D_\mathrm{it}\frac{\partial (1-f_s)}{\partial E}dE_t
\end{eqnarray} valid for positively charged donor trap states, where $f_s$ is the Fermi-Dirac distribution function \cite{sze2006}.

\section{Band to band tunneling model}
From Kane's model for tunneling \cite{kane1961theory}
\begin{eqnarray}\label{eq_btbt}
I_{BTBT} = AWV_R\Big(\frac{|\mathcal{E}|}{\mathcal{E}_0}^P\Big)e^{-\frac{B}{\mathcal{E}}}
\end{eqnarray}where $A$, $B$, $P$ are material fitting parameters. $V_R$ is the effective bias controlled by both gate and drain biases by $V_R= F_\mathrm{sat} E_\mathrm{TW}/q$. $F_\mathrm{sat}$ (shown below) captures the superlinearity in $I_\mathrm{DS}-V_\mathrm{DS}$ and controls $V_R$ when $V_\mathrm{DS}$ is lower than the tunneling energy window, $E_\mathrm{TW}$. From the Landauer formalism and using $F_\mathrm{sat}$ = $I_\mathrm{DS}(V_\mathrm{DS})/I_\mathrm{DSmax}$, 
\begin{eqnarray}\label{fsat1}
F_\mathrm{sat} = \frac{k_BT}{\mu_s-E_\mathrm{cch}} \log\bigg[\frac{\mathrm{exp}\Big(\frac{E_\mathrm{vs}}{k_BT}\Big)+\mathrm{exp}\Big(\frac{\mu_s-qV_\mathrm{DS}}{k_BT}\Big)}{\mathrm{exp}\Big(\frac{E_\mathrm{vs}}{k_BT}\Big)+\mathrm{exp}\Big(\frac{\mu_s}{k_BT}\Big)}{\times}...\nonumber\\\frac{\mathrm{exp}\Big(\frac{E_\mathrm{cch}}{k_BT}\Big)+\mathrm{exp}\Big(\frac{\mu_s}{k_BT}\Big)}{\mathrm{exp}\Big(\frac{E_\mathrm{cch}}{k_BT}\Big)+\mathrm{exp}\Big(\frac{\mu_s-qV_\mathrm{DS}}{k_BT}\Big)}\bigg]\nonumber\\
\end{eqnarray}where $\mu_s$ is the source Fermi energy, and $E_\mathrm{vs}$ and $E_\mathrm{cch}$ are the source valence and channel conduction band edges respectively. The above expression works for both degenerately and non-degenerately doped sources and for all gate voltages. 
\begin{figure}
\centering
\includegraphics[width=2in]{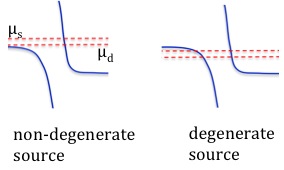}
\subfigure{\includegraphics[width=2.7in]{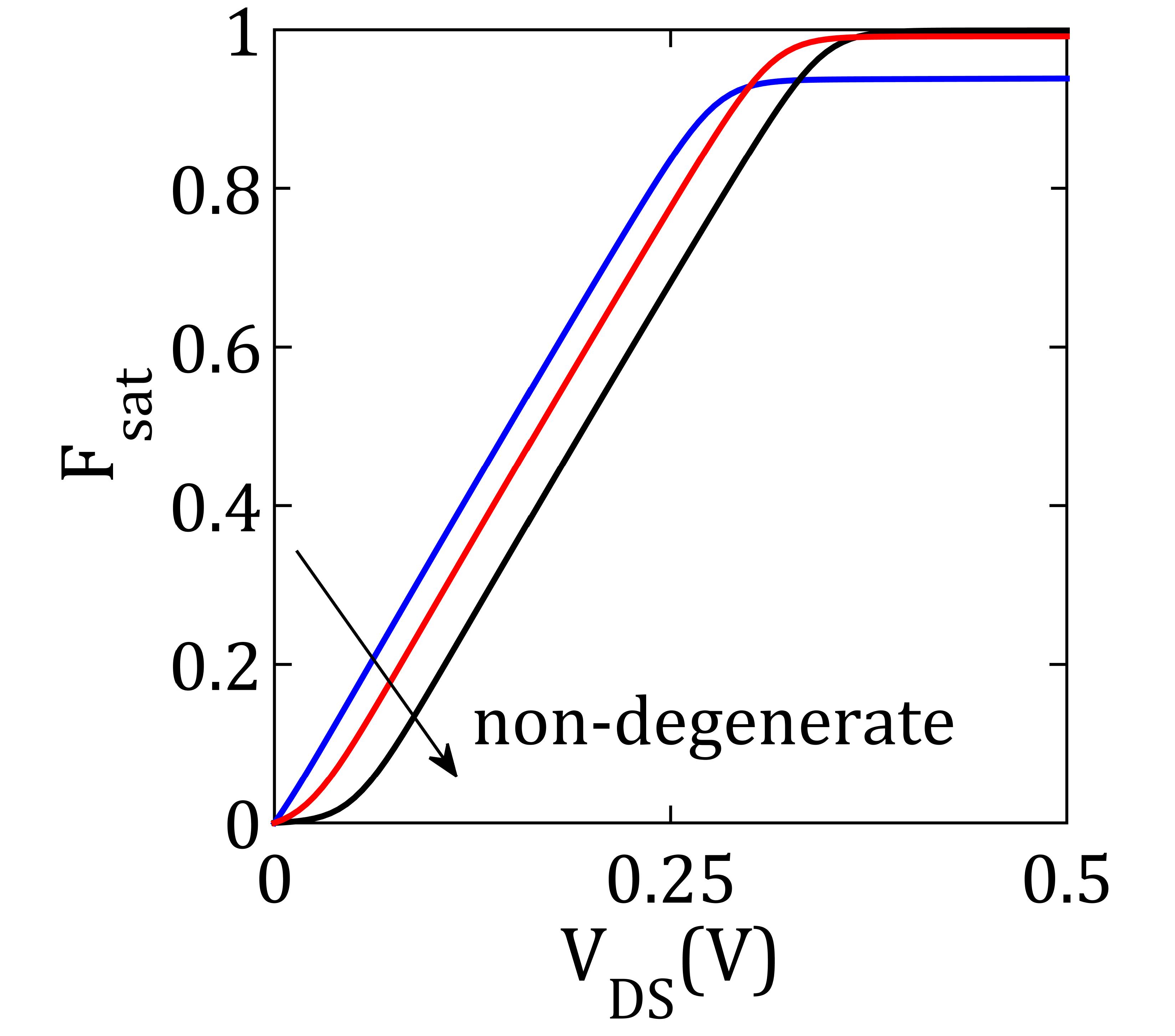}}
\caption{(a) Impact of source degeneracy on the saturation function. Non-degenerate source requires a minimum drain voltage to trigger current (drain threshold voltage) and therefore produces a super-linear output characteristics. This mechanism of non-linearity is captured through $F_\mathrm{sat}$ (Eq. \ref{fsat1})}
\label{fsat}
\end{figure}
For a non-degenerate source ($E_\mathrm{vs}-\mu_s<0$), the drain current does not initiate until the Fermi window ($\mu_\mathrm{s}-\mu_\mathrm{d}$) is large enough to penetrate the valence band of the source. As a result the output current is non-linear for small drain bias (black curve in Fig. \ref{fsat}). For degenerate source, the Fermi window is already inside the tunnel window (at zero drain bias) and therefore the output current increases linearly. Transport for all such conditions is captured with the above $F_\mathrm{sat}$ function. 

\begin{figure}
\centering
\includegraphics[width=2.9in]{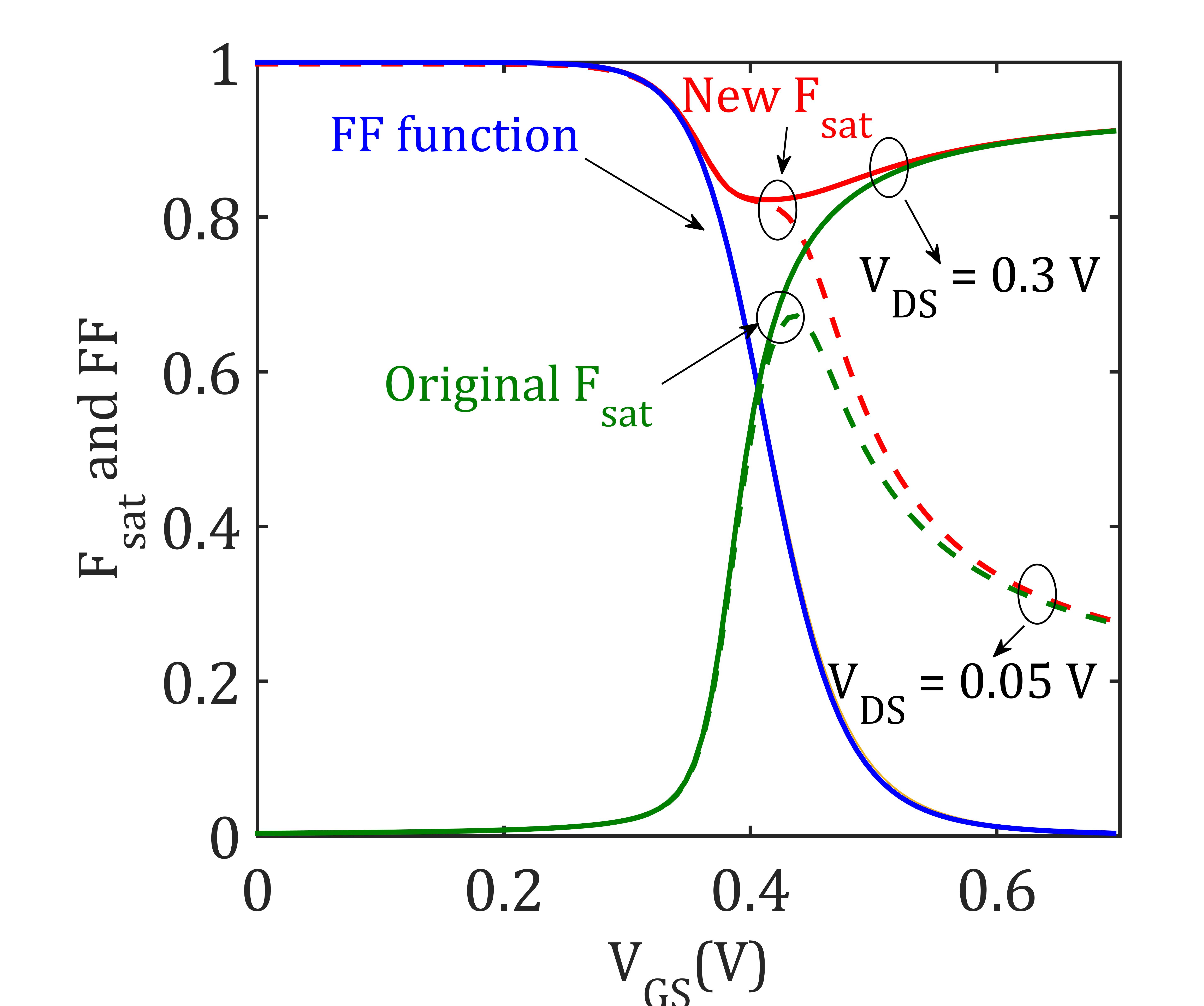}
\caption{Below the threshold voltage, $V_t$, $F_\mathrm{sat}$ in Eq. \ref{fsat1} yields incorrectly low values. The $FF$ function, as shown in Eq. \ref{eq:fsat2}, fixes this problem.}
\label{fsat-combined}
\end{figure}

However,  the denominator, which is a normalizing factor, $\mu_s-E_\mathrm{cch}$ in the expression of $F_\mathrm{sat}$ is valid for above threshold only ($V_\mathrm{GS}>V_t$) and therefore for $V_\mathrm{GS}<V_t$, the above expression of $F_\mathrm{sat}$ predicts an incorrectly low value. Below $V_t$, the tunnel window is extremely small and therefore $F_\mathrm{sat}$ should be 1 for all practical purposes. We solve this problem by introducing a function $FF$ which is 1 below threshold and 0 above threshold and take the weighted average as below, 
\begin{eqnarray}\label{eq:fsat2}
F_\mathrm{sat,combined} &=& FF+(1-FF)\times F_\mathrm{sat}\\
FF &=& \frac{1-e^{-V_\mathrm{DS}/k_BT}}{1+e^{\frac{\psi-V_t}{U_t}}}
\end{eqnarray}where $U_t$ is of the order of a few $k_BT$, fitted to get the correct transition. The combined function along with the original $F_\mathrm{sat}$ is shown in Fig. \ref{fsat-combined}. 

The tunnel energy window can be found from
\begin{eqnarray}\label{eq_win}
E_\mathrm{TW} = U_0\log\Big[1+e^{\frac{\psi-V_t}{U_0}}\Big]
\end{eqnarray}

$U_0$ is a parameter that defines the sharpness of the band edge (Urbach tail times the geometric gate efficiency). In principle, $U_0$ should be temperature dependent, however previous studies on Urbach tail reveals weak temperature dependence \cite{johnson1995,greeff1995}. We use $U_0 = K_BT_0/ \gamma_t$, where $T_0 = 300$ K, and $\gamma_t = \gamma(T_0/T)^\beta$, where $\gamma$ and $\beta$ are fitting parameters. 
\begin{figure}
\centering
\includegraphics[width=1in]{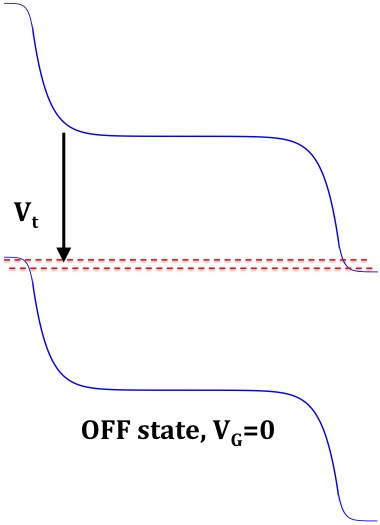}
\subfigure{\includegraphics[width=1in]{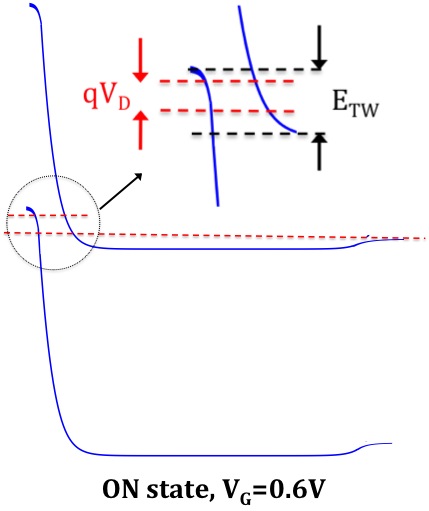}}
\subfigure{\includegraphics[width=2.5in]{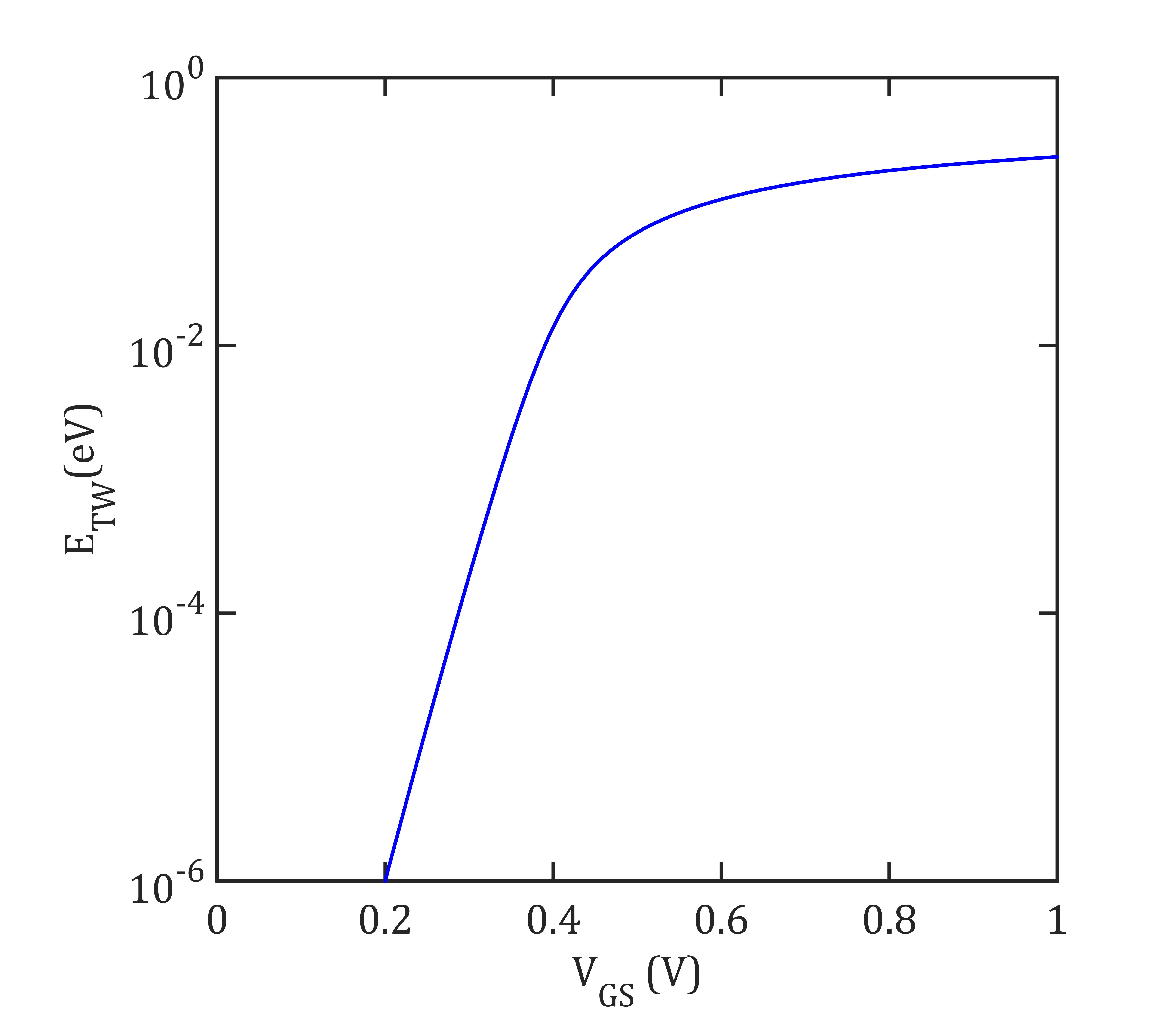}}
\caption{As the TFET goes from OFF to ON state, the energy tunnel window $E_\mathrm{TW}$ changes from an exponential function (Urbach-tail-limited) to a linear function at voltages above threshold, $V_t$.}
\label{etw}
\end{figure}
Before the bands are overlapped (at $V_\mathrm{GS} = V_t$), it can be assumed that the tunnel window increases exponentially, instead of an abrupt turn-ON (at $V_\mathrm{GS} = V_t$) at , as a result of the exponentially decaying states above the valence band in the source (for an n-channel TFET). Above $V_t$, the window increases linearly, which is captured by Eq. \ref{eq_win}, as expected (Fig. \ref{etw}). For a material with low quantum capacitance (such as the III-V semiconductors), it is possible to have a tunnel window $E_\mathrm{TW}$ larger than $V_\mathrm{DS}$ (Fig. \ref{etw}). From low to high bias, $F_\mathrm{sat}$ obtains the correct fraction of the $E_\mathrm{TW}$ that contributes to current (through $V_R$ in Eq. \ref{eq_btbt}). 
With negative drain bias, the surface potential decreases due to increased quantum capacitance resulting in decrease in the tunnel window and negative differential resistance. 

\begin{figure}
\centering
\includegraphics[width=3in]{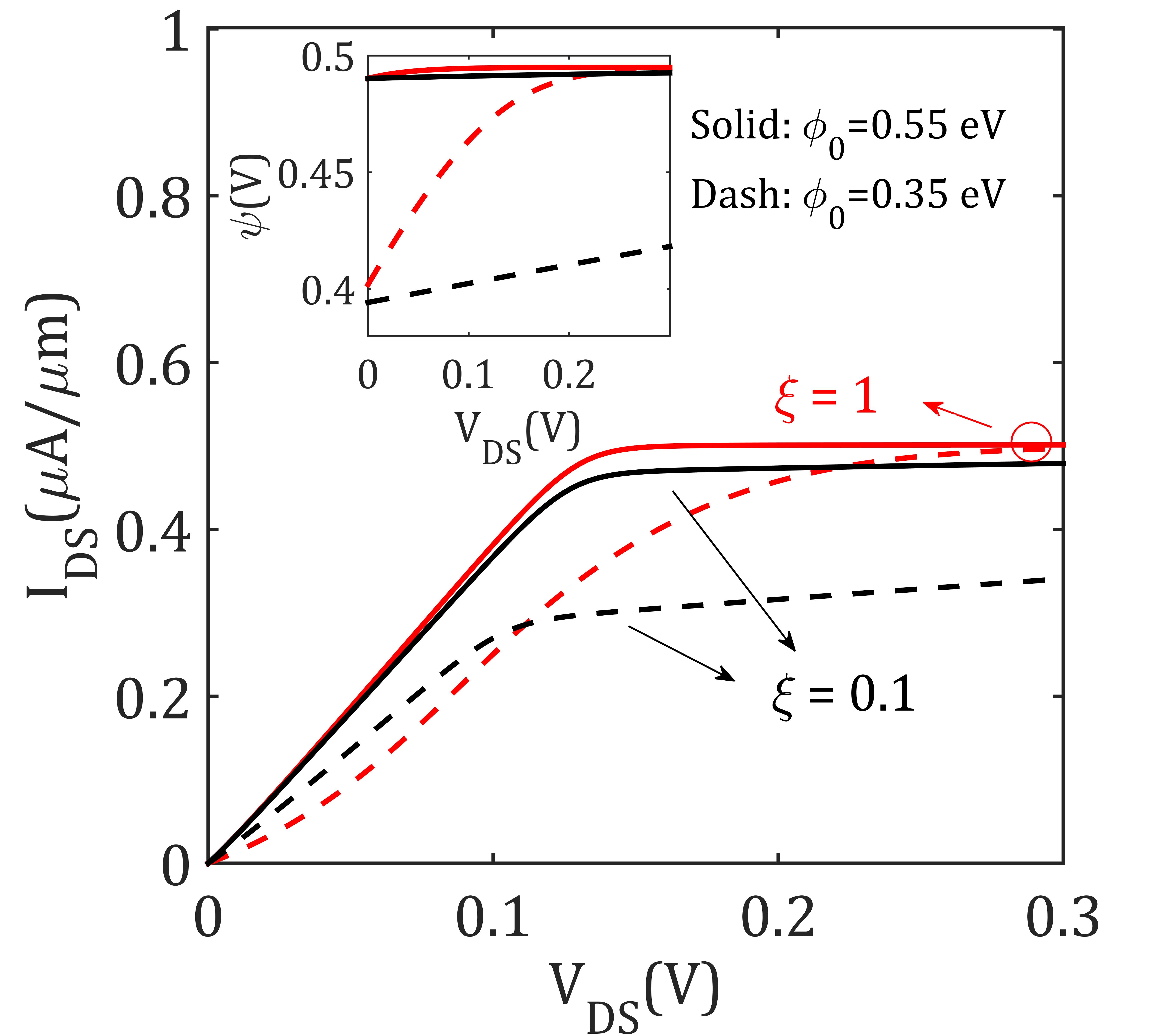}
\caption{Impact of the quantum capacitance parameters on the output characteristics. When $\psi$ at low bias is low (low $\phi_0$), a high value of $\xi$ makes the surface potential very sensitive to the drain bias. Therefore, it leads to a non-linear output characteristics. For low values of $\xi$, $\phi_0$ changes the surface potential (Eq. \ref{eq:psi}) and the energy tunnel window (Eq. \ref{eq_win}) substantially, leading to the change in ON current. }
\label{phi-id-vd}
\end{figure}

\begin{figure}
\centering
\includegraphics[width=2.9in]{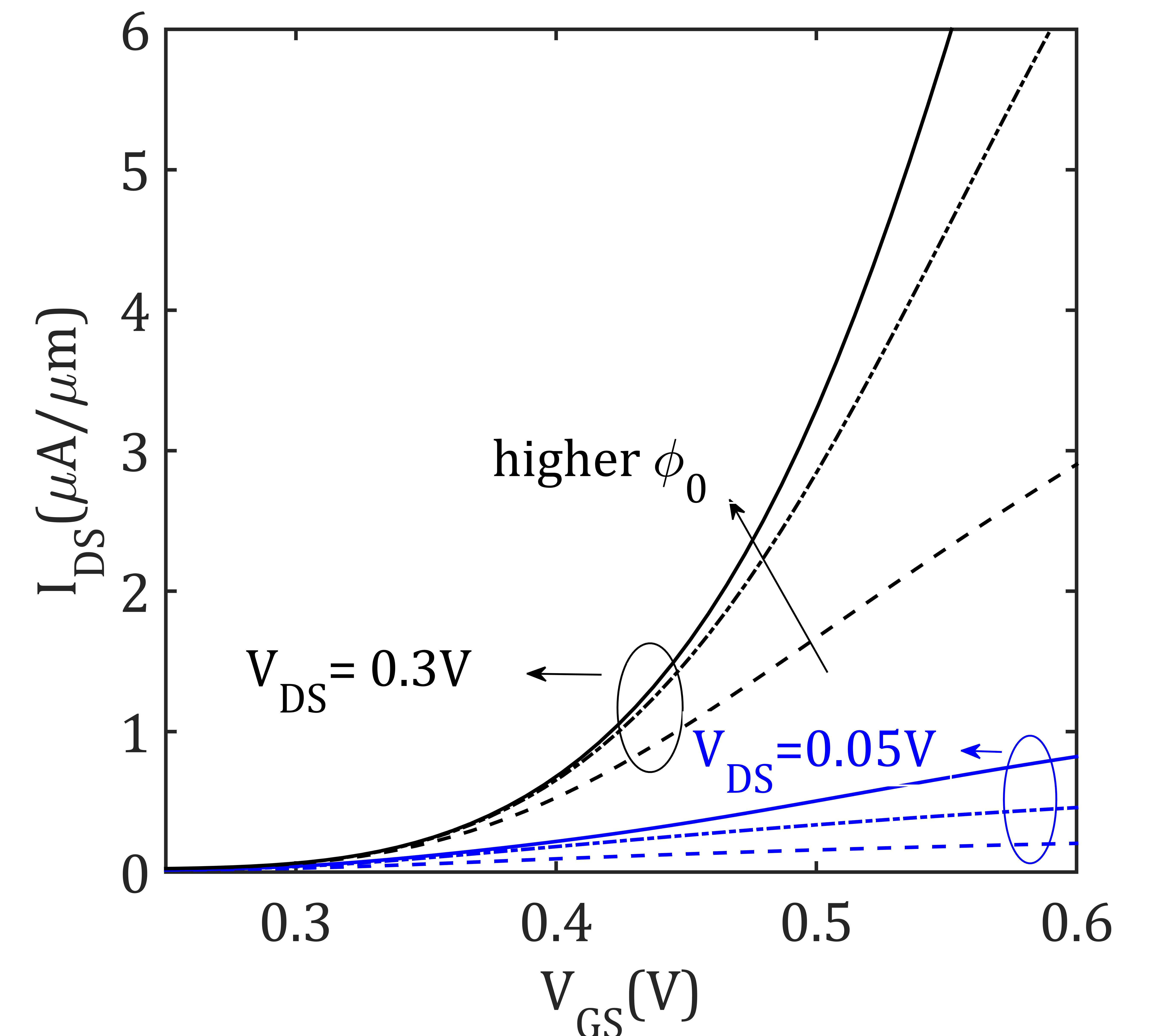}
\caption{(a) Transconductance varies strongly with $\phi_0$, since it affects the tunnel window and the ON current. Smaller values of $\phi_0$ (for small drain bias) pins the surface potential (for $V_\mathrm{GS}>>\phi$, $\psi \approx \phi = \phi_0+\xi V_\mathrm{DS} \approx \phi_0$, as discussed in the text) and the current saturates at low values. }
\label{gm}
\end{figure}

Once we know the TAT and BTBT contributions from Eqs. \ref{compact_current} and \ref{eq_btbt}, we add the two components to find the total current. Figs. \ref{phi-id-vd} and \ref{gm} show the impact of some of the fitting parameters on the current-voltage characteristics. Together with $\xi$, $\phi_0$ controls the super-linearity of the output characteristics and the maximum ON current. The output characteristics is shown in Fig. \ref{phi-id-vd} for different values of $\phi_0$ and $\xi$. When $\xi$ = 1, a smaller value of $\phi_0$ leads to super-linear output characteristics since $\psi$  changes strongly with drain bias (inset). This is a result of TFET’s inherent DIBL and another mechanism, in addition to the the lack of source degeneracy (discussed earlier), that may cause super-linear output characteristics. For materials with high density of states (such as silicon), $\phi_0$ will be lower, resulting in stronger super-linear behavior. The super-linearity can diminish if $\phi_0$ is already high at small drain bias and therefore $\psi$  changes little as bias is increased. The super-linear behavior can also diminish if the value of $\xi$ is sufficiently low ($\xi = 0.1$ in Fig. \ref{phi-id-vd}), regardless of $\phi_0$. However in this case $\phi_0$ strongly influences the energy tunnel window (Eqs. \ref{eq:psi}, \ref{eq_win}); above $V_t$, $E_\mathrm{TW, high V_\mathrm{DS}} =  \psi_\mathrm{high V_\mathrm{DS}}-V_t$, and therefore change of $\phi_0$ changes the ON current. 


Since $\phi_0$ changes the energy tunnel window and the current, the transconductance is also dependent upon this parameter. Fig. \ref{gm} shows the transfer characteristics for various $\phi_0$ values and two different drain biases. Higher values of $\phi_0$ lead to higher transconductance and the low drain bias transconductance is always lower than the higher bias. This is  because of the higher quantum capacitance at low drain bias and the slow increase of $\psi$ with $V_\mathrm{GS}$ (Fig. \ref{drain_control}). Since the carrier density is a function of temperature, the parameter $\phi_0$ is also temperature dependent.

\begin{figure}
\centering
\includegraphics[width=3.4in]{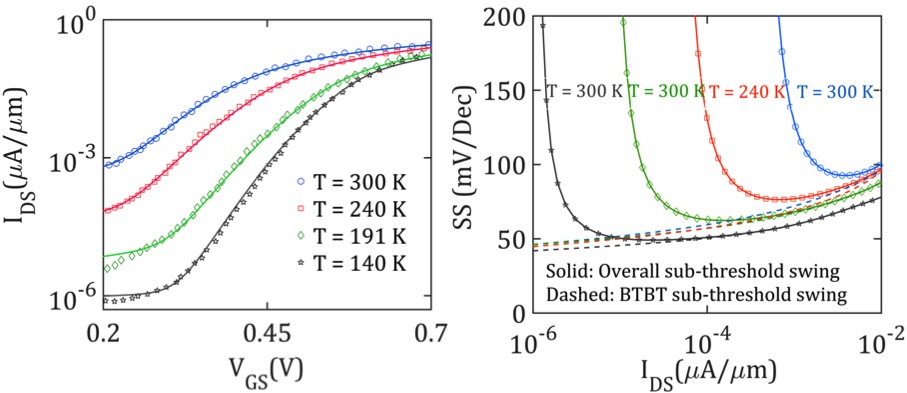}
\caption{Model fit (line) of the experimental data (symbols) in Ref. \cite{tao2015} - a InGaAs/GaAsSb heterostructure TFET. The model yields good fit at different temperatures with parameters shown in Tables I and II.}
\label{tao-fit}
\end{figure}
\begin{table}
\caption{Temperature independent parameters used in Fig. \ref{tao-fit}}
\centering
\begin{tabular}{|c|c|c|}
\hline
Symbol & Value \\    \hline
$E_g$ & 0.7 eV \\    \hline
$m^*$ & 0.041m\\ \hline
$t_\mathrm{ox}$ (EOT) & 1 nm\\ \hline
$t_\mathrm{semi}$ & 31 nm\\ \hline
$\xi$ & 0.05 \\ \hline
$deg$ & 15 meV (degenerate)\\ \hline
$D_\mathrm{it}$ & $3\times10^{15}\mathrm{/cm}^2$-eV\\ \hline
$\beta$ & 0.03\\ \hline
$\Delta E $ & 0.194 eV\\ \hline
\end{tabular}
\end{table}
\begin{table}
\caption{Temperature dependent parameters used in Fig. \ref{tao-fit}}
\centering
\begin{tabular}{|c|c|c|c|c|c|}
\hline
T (K) & v$_\mathrm{shift} (V)$ & $\phi_0$ (V) & $\zeta$ & $\gamma$ & $\sigma$ \\   
 & & & & & ($\times10^{-17}/\mathrm{cm}^2$)\\ \hline
300 & 0.07 & 0.04 & 0.25 & 0.7 & 5 \\    \hline
240 & 0.02 & 0.04 & 0.15 & 0.7 & 2.5\\ \hline
191 & -0.025 & 0.04 & 0.08 & 0.9 & 2 \\ \hline
140 & -0.11 & 0.09 & 0.08 & 0.9 & 2\\ \hline
\end{tabular}
\end{table}

\section{Model fit of experimental data}  

Fig. \ref{tao-fit} shows the experimental transfer characteristics along with the model fits for an InGaAs/GaAsSb quantum well vertical TFET \cite{tao2015}. The fitting procedure is summarized as following. The bandgap, effective mass and carrier capture cross-section due to the traps are adopted from the literature for the particular channel material (Table I). The TAT parameters ($d_\mathrm{gen}$, $\Delta E$, $D_\mathrm{it}$) are varied to fit the leakage current of the device at different temperatures. The extracted interface trap density is found to be $D_\mathrm{it} = 3\times10^{11}/\mathrm{cm}^2$-$\mathrm{eV}$, which signifies the interface trap density between the GaAsSb source and InGaAs channel. The interface between the two epitaxially grown structures defect density is lower than the oxide - III-V interface. The intrinsic band steepness parameters ($\gamma$, $\beta$) are then varied to match the sub-threshold swing at different temperatures. The electrostatic parameters ($t_\mathrm{semi}$, $\xi$, $\zeta$, $\phi_0$) are tuned to match the current magnitudes in the ON state and the transconductance. The model parameters to fit the data shown in Fig. \ref{tao-fit} are shown in Tables I and II. 

With the extracted BTBT parameters ($\gamma$, $\beta$), we extract the intrinsic sub-threshold swing as a function of temperature as shown in Fig. \ref{tao-fit}c. As discussed earlier, the TAT contribution obscures the intrinsic, steep BTBT as evident from Fig. \ref{tao-fit}. The intrinsic BTBT is found to be around 50 mV/dec (in the sub-threshold regime, below $10^{-4}$ $\mu$A$/\mu$m), which can be considered as the product of the Urbach tail and the geometric gate efficiency. Reported values of Urbach tail are around 30 mV/dec, therefore our estimate of the geometric gate efficiency is $\sim$0.6. The low gate efficiency can be justified by the particular device structure, where the source-channel interface is deliberately placed far from the gate oxide-channel interface (with the InP cap) to minimize the impact of oxide interface traps. 

\begin{figure}
\centering
\includegraphics[width=3.5in]{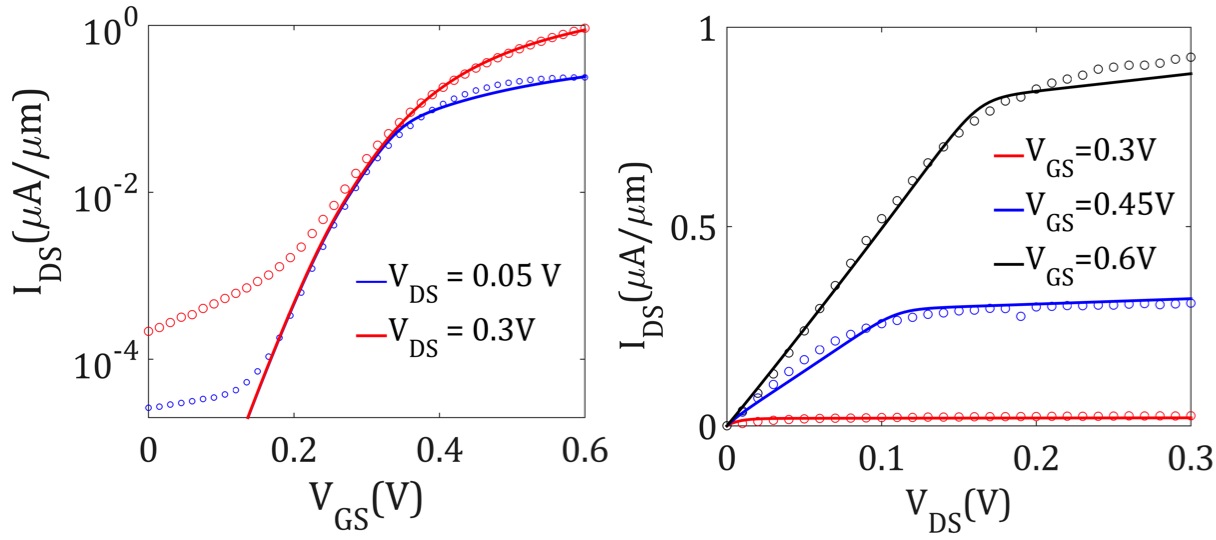}
\caption{Simultaneous fitting of $I_\mathrm{DS}$-$V_\mathrm{GS}$ and $I_\mathrm{DS}$-$V_\mathrm{DS}$, symbols are experimental data from Ref. \cite{tao2015}.}
\label{tao-fit2}
\end{figure}

In order to make sure the model can simultaneously fit both transfer and output characteristics, we have fitted data at $T = 94$ K as shown in Fig. \ref{tao-fit2}. The model deviates from the data for this particular device for low $V_\mathrm{G}$ values due to possible other leakage paths that were not included in the model. 
\section{Verilog-A model and circuit simulation}
In this section, we demonstrate a Verilog-A implementation of our compact model in Advanced Design System (ADS) circuit simulator. 

\begin{figure}
\centering
\includegraphics[width=1.8in]{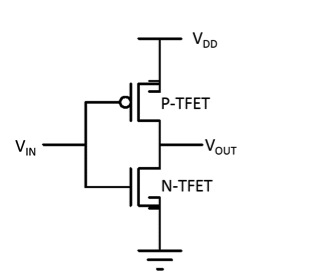}
\subfigure{\includegraphics[width=3.3in]{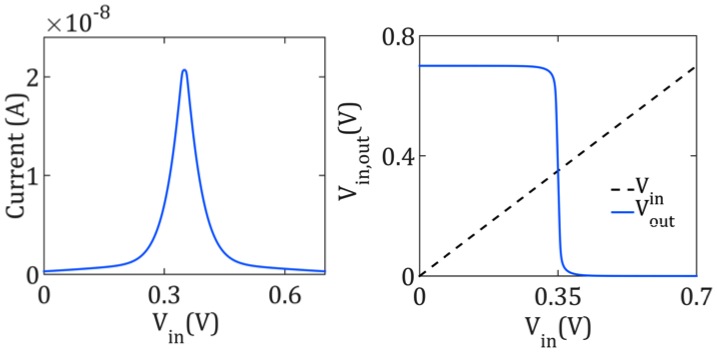}}
\caption{TFET-based inverter simulation which uses the developed Verilog-A model fitted against measurements. We show the quasi-static inverter transfer characteristics along with the supply current vs. input voltage. }
\label{cmosi}
\end{figure}

In Fig.\ref{cmosi}, we show TFET-based inverters using the N-TFET devices and symmetric P-TFETs. We implement the model described in this work in Verilog-A and calibrate against measured DC-currents shown in Fig.\ref{tao-fit2}, along with capacitances. We perform the inverter simulations in ADS. The resulting voltage-transfer-characteristics of the inverter are shown in the figure along with terminal current as a function of input voltage. The simulations show the convergence robustness of the Verilog-A model.

\begin{figure}
\centering
\includegraphics[width=2.8in]{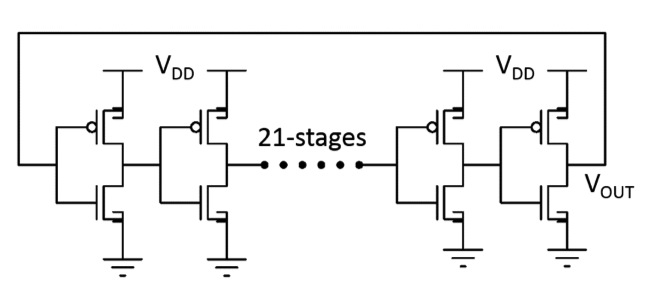}
\subfigure{\includegraphics[width=2.5in]{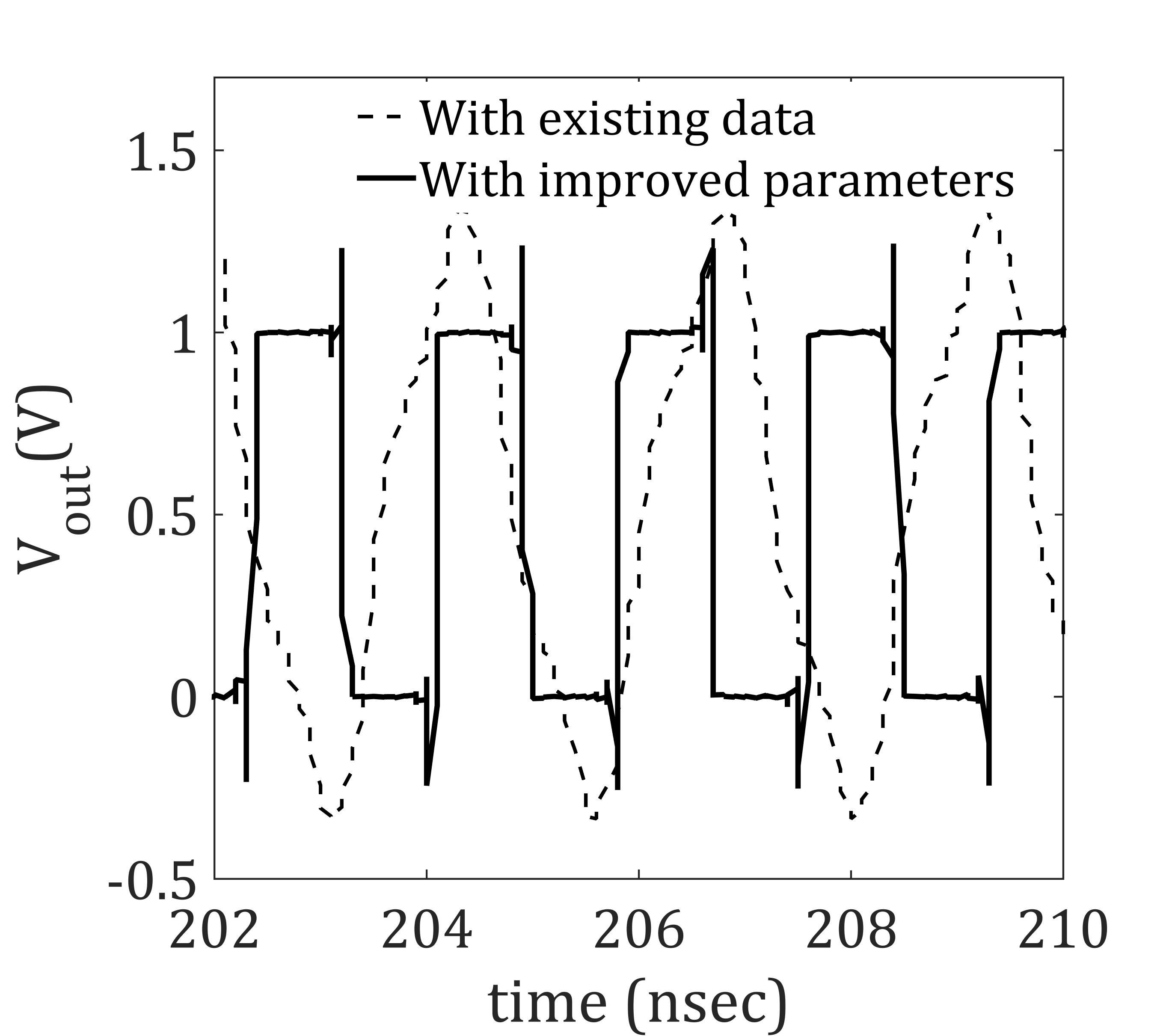}}
\caption{We simulate a 21-stage Ring Oscillator (RO) using the model fitted to the expermental data that is used as baseline and the model whose parameters are projected for improved device performance. The output time-domain waveforms show the oscillation frequency which improves with better device design compared to the baseline. The frequency boost is about 30$\%$. }
\label{ro}
\end{figure}
The inverter is used as a building block to simulate a 21-stage ring oscillator (RO) whose output  voltage waveforms are shown in Fig.\ref{ro}. TFET shown in Fig. \ref{tao-fit2}, the baseline TFET at T = 300 K, shows oscillation frequency of about 500 MHz at $V_\mathrm{DS}$ = 1 V. The low frequency (or transit-time) in the baseline TFET is because of low ON-current and high OFF-current (limited by the TAT). Better electrostatic design along with a reduction in $D_\mathrm{it}$ yields improved SS. Use of channel material that can provide lower bandgap and carrier-effective mass can boost ON-current. The model that incorporates these improvements at the device-level can yield a reduction in transit-time and an increase of 30$\%$ in oscillation frequency (solid line in Fig. \ref{ro}). The Verilog-A model developed in this work can thus provide a tool-guide for innovative device design for desired system performance.

\section{Conclusion} We present a compact model with the inclusion of trap assisted tunneling allowing studies of surface trap effects on TFET circuit performance. The model captures the structural and material parameters that influence TAT. We provide compact expressions of channel potential and electric field that accurately capture the TFET quantum capacitance. The model can also be used to extract intrinsic sub-threshold swing from a given experimental data. We implement the model in Verilog-A and present simulation of simple circuits based on TFETs. The model for the first time evaluates the impact of non-idealities on TFET circuit performance. 

\section*{Acknowledgment}
This work was supported by NSF - E3S center, Award 0939514, and the NCN-NEEDS Program, Grant 1227020-EEC, supported by Semiconductor Research Corporation. 

\bibliographystyle{apsrev4-1}
%

\end{document}